\documentclass[%
 reprint,
showpacs,%preprintnumbers,https://www.overleaf.com/project/5e5e33ba1e493b000149ee19
 amsmath,amssymb,amsfonts,amsthm,amscd,bbm,
 aps,
prb,
]{revtex4-1}

\usepackage{graphicx}% Include figure files
\usepackage{dcolumn}% Align table columns on decimal point
\usepackage{bm}% bold math
\usepackage{float}
\usepackage{placeins}
\usepackage{lipsum}
\usepackage{color}
\definecolor{Blue}{rgb}{0.3,0.3,0.9}
\definecolor{Red}{rgb}{0.9,0.3,0.3}
\definecolor{Green}{rgb}{0.3,0.6,0.3}

\newif\ifNOSUP \NOSUPtrue
\begin{document}

%\preprint{JO/RAR-QHHF}

%\linenumbers

\title{Revision of the edge channel picture for the integer quantum Hall effect}% Force line breaks with \\

\author{Josef Oswald}
\email{Josef.Oswald@unileoben.ac.at}
%\altaffiliation[Also at ]{Physics Institute, University of Leoben, Austria}%Lines break automatically or can be forced with \\
\affiliation{%
Institut f\"{u}r Physik, Montanuniversit\"{a}t Leoben, Franz-Josef-Strasse 18, 8700 Leoben, Austria
}%

\date{$Revision: 1.0 $, compiled \today}

%%%%%%%%%%%%%%%%%%%%%%%%%%%%%%%%%%%%%%%%%%%%%%%%%%%%%%%%%%%%%%%%%%%%
\begin{abstract}

State of the art computing opens now a new window to the integer quantum Hall effect (IQHE) regime, which enforces a major revision of the common knowledge accumulated so far. By our record-breaking application of the Hartree-Fock method we use up to 3000 electrons distributed over up to 5000 states for almost macroscopic system size of 1000x1000nm. In particular, the formation of compressible and in-compressible edge stripes turns out to develop essentially different from the common picture used so far. Oppositely to the theory Chklovskii, Shklovskii and Glazman (CSG), the narrow channels, as assumed by the early models of the IQHE, do not widen up into wide compressible stripes. Instead, the wide compressible stripes of CSG transform into a mixture of clusters of full and empty spin-split LLs, while the cluster boundaries create a network of still narrow quantum channels sitting on top of the wide compressible stripes. On this background the early models based on narrow edge channels do not suffer from neglecting electron-electron interaction as falsely stated in the past. Quite oppositely, in contrast to the common believe, our modelling demonstrates that also the IQHE regime carries the hallmark of many-body physics which stabilizes narrow edge channels also in the presence of electron-electron interaction.

\end{abstract}



\maketitle

%\tableofcontents

%%%%%%%%%%%%%%%%%%%%%%%%%%%%%%%%%%%%%%%%%%%%%%%%%%%%%%%%%%%%%%%%%%%%

%\section{\label{sec:level1}Introduction}
\section{Introduction}

Almost 40 years after discovery, the quantum Hall effect \cite{KliDP80} was formally included among the select group of high-precision experiments to form the basis of a new SI system. In his essay to celebrate this achievement \cite{vonKlitzing2019}, Klaus von Klitzing also points out "that a microscopic picture of the quantum Hall effect for real devices with electrical contacts and finite current flow is still missing." A fundamental question in this context is the microscopic nature of compressible stripes, which is exactly the topic of this paper. Looking back to the 1980-ties and 1990-ties there can be identified three major achievements for the theory of the IQHE. On the one hand there is the scaling theory (see review of B. Huckestein\cite{Huckestein1995ScalingEffect}) based on a network of narrow quantum channels, on the other hand Chklovskii, Shklovskii and Glazman\cite{Chklovskii1992} (CSG) have shown that such narrow channels cannot exist and transform into wide compressible stripes (CS) if taking into account electron-electron (e-e) interaction (on the single particle level). The third achievement is the theoretical approach of the bubble and stripe phases in ultra clean systems\cite{Fogler1996d, Koulakov1995}. In this paper, we focus on the theoretical foundations of narrow quantum (edge) channels, as used e.g. by the scaling theory and the wide compressible (edge) stripes, as proposed by the theory of CSG\cite{Chklovskii1992}. Both have been successfully justified many times\cite{RN196, RN357, Karmakar2015, RN330, RN212, Panos2014CurrentInvestigation, Weis2011, PasRIE14, KenSKO17} although they rest on conflicting theoretical foundations\cite{Werner2020SizeRegime}. In recent papers\cite{Romer2021TheRegime, OswaldPRB2017, OswR17, Oswald2020, Werner2020SizeRegime} we argue that many body interactions play an important role also for the IQHE regime. Werner et al. have shown that within a system of size $1000 \times 1000nm$ many-body interactions are almost fully-fledged \cite{Werner2020SizeRegime}. Addressing a macroscopic system beyond that size by a single Hartree-Fock (HF) setup would exceed any computing capabilities so far. Werner et al. suggest HF modelling of macroscopic systems beyond $1000 \times 1000nm$ by joining several plaquetts of size $1000 \times 1000nm$, while addressing each plaquett by HF individually.

In this paper we demonstrate for the first time and explain explicitly how narrow edge channels survive the e-e interaction and the formation of wide compressible edge stripes. We show that this is possible only on the basis of many-body interactions. As a first step towards following the suggestion of Werner et al., we use a single plaquette of $1000 \times 1000nm$ size in order to model the soft edge potential region of a QH sample.  

\section{Method}

We use a Hartree-Fock approach as explained in previous work\cite{SohR07, OswR17,OswaldPRB2017,Oswald2020}, where we have shown that the key-mechanism dictating the electron system is a Hund's rule behaviour for the occupation of the spin-split Landau levels (LL). The resulting $g$-factor enhancement appears as an almost local quantity depending on the local filling factor. We argue that the lateral variation of the $g$-factor enhancement due to then lateral filling factor variation and hence, the laterally varying Zeeman energy has to be considered in addition to the laterally varying Hartree potential, leading to a modified effective total potential for an effective single electron picture\cite{OswaldPRB2017}. This strongly modifies the screening behaviour and can even result in an effective overall attractive potential (see Appendix A). It creates bubbles and stripes in ultra clean samples\cite{Oswald2020}, or it results in the formation of clusters of full and empty spin-split LLs which's arrangement is triggered by the disorder or edge potential \cite{Romer2021TheRegime,OswR17,OswaldPRB2017}. The boundaries of those clusters finally create narrow channels that align along the edge or along the random potential fluctuations.

\begin{figure*}[tb]
%\mbox{ }  \hfill Hartree \hfill non-interacting \hfill  \mbox{ }\\
%\hspace*{-0.8\columnwidth} $\uparrow$ \hspace{0.8\columnwidth}$\downarrow$
(a) \includegraphics[width=0.45\textwidth,clip=true,trim=120 30 120 0]{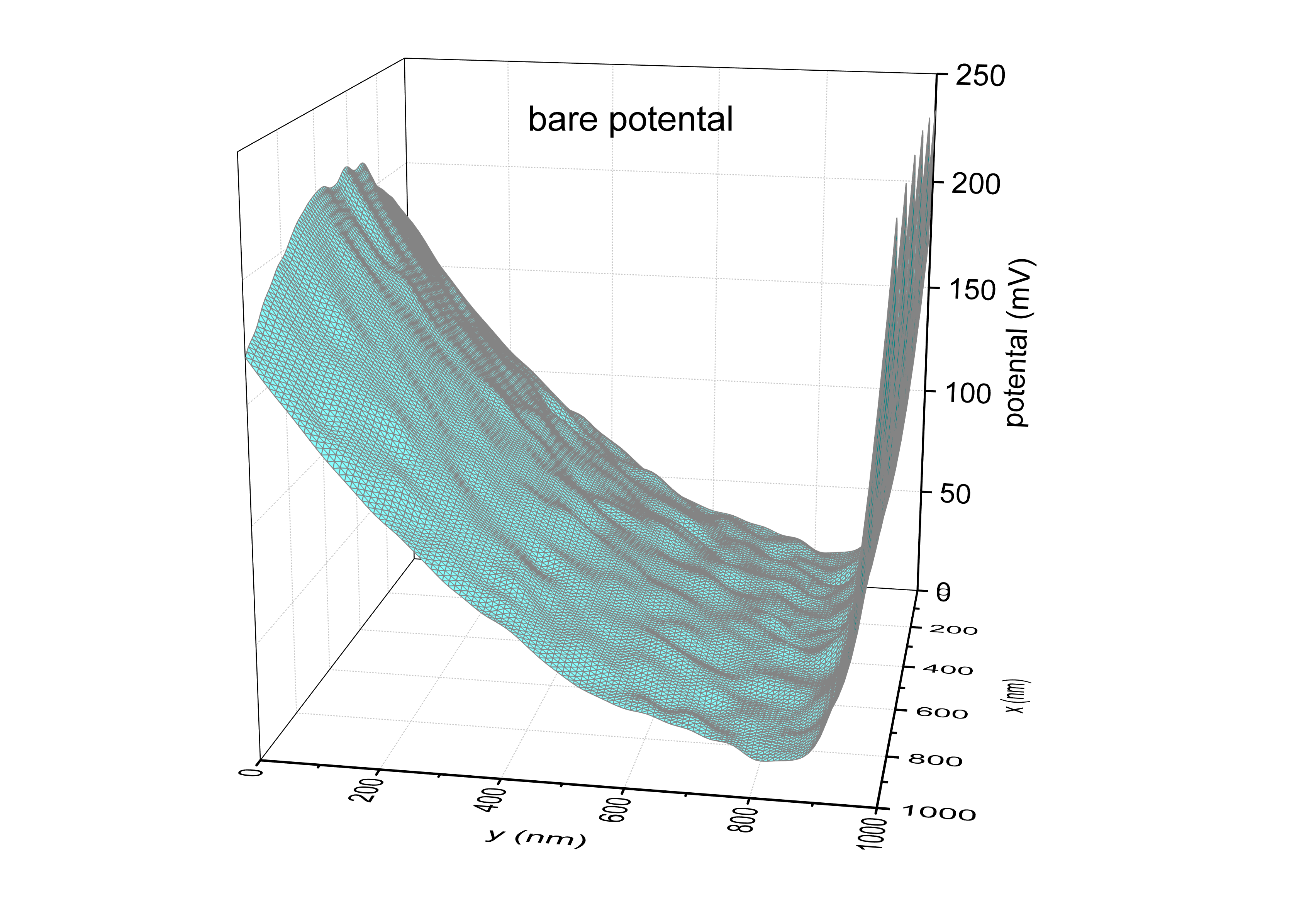}
(b) \includegraphics[width=0.45\textwidth,clip=true,trim=120 30 120 0]{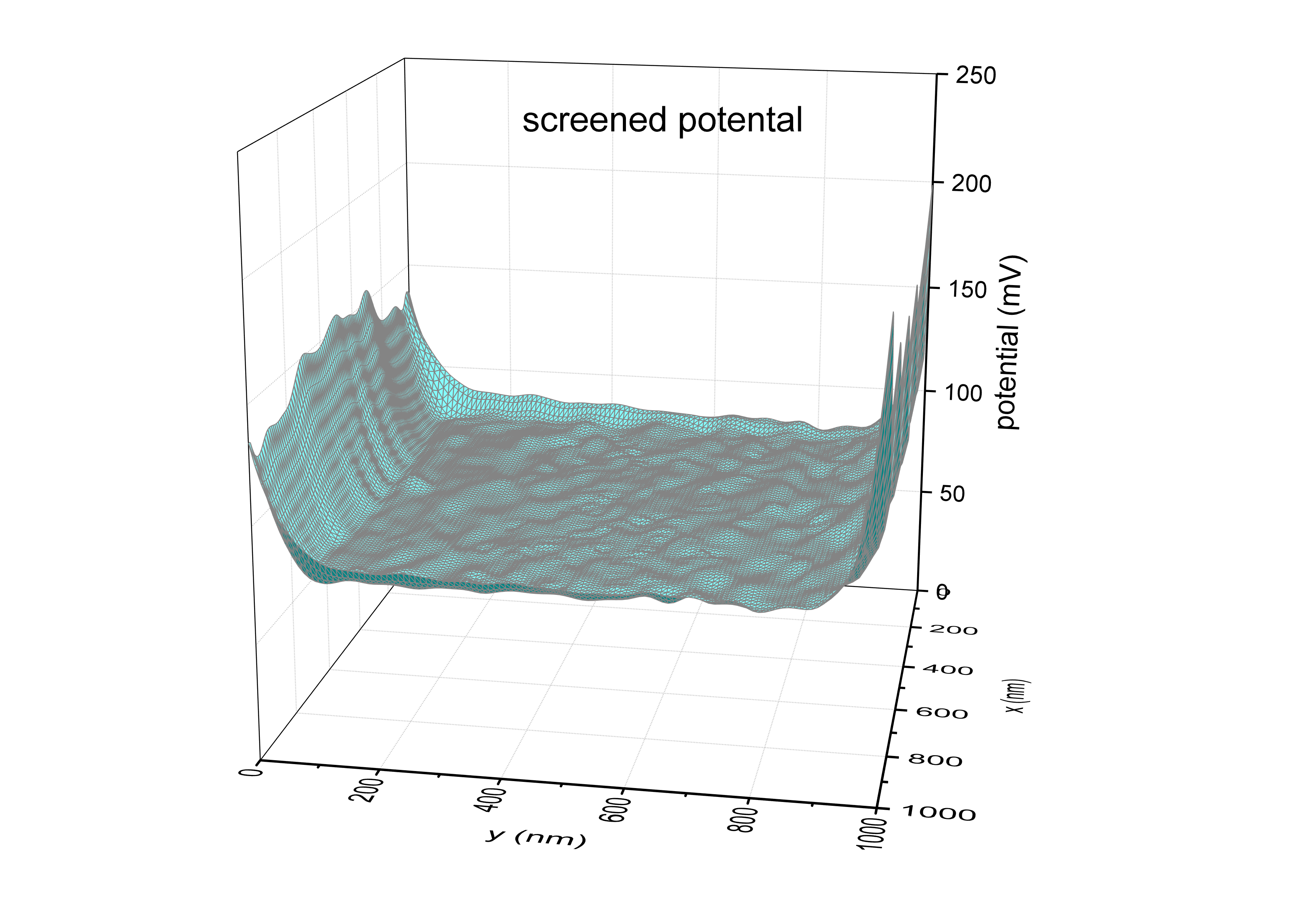}\\
(c) \includegraphics[width=0.65\textwidth,clip=true,trim=000 70 00 0]{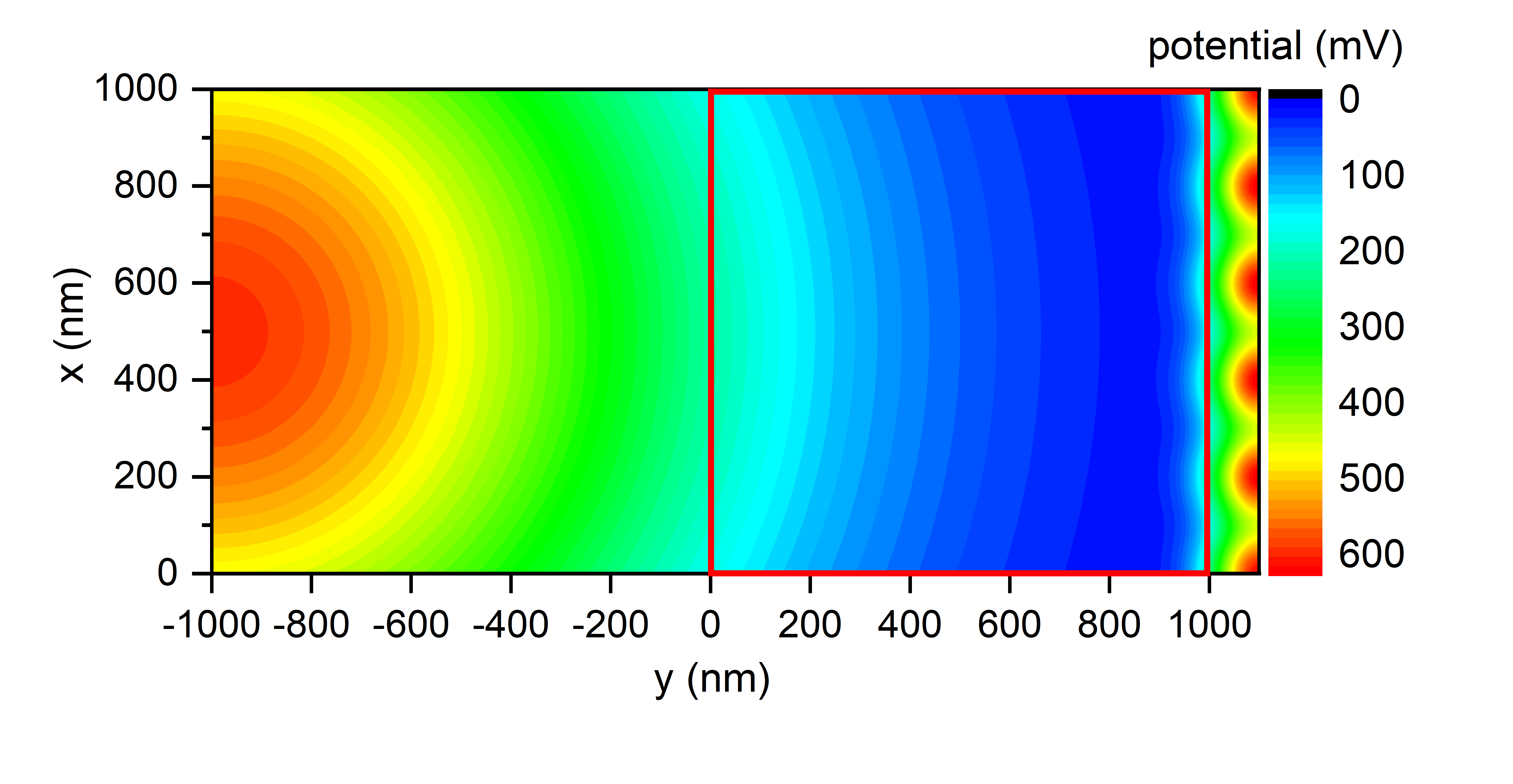}
%(d) $\downarrow$\includegraphics[width=0.45\textwidth]{CD_nu_450S1-00_.png}
\caption{
\label{fig_pot}
(a) Bare potential as used for the HF simulation as input. The superimposed bare random potential has a fluctuation amplitude of about $\pm 5mV$. (b) Screened potential as obtained from the HF solution at a magnetic field $B = 2.5 T$ and an average carrier density $n = 1.4\cdot 10^{11} cm^{-2}$. The superimposed random potential keeps its strength while the slope of the edge potential is completely flattened by screening. (c) Illustration of how the bare potential is composed by the superposition of a number of Gauss peaks as explained in text. The actual potential used in the HF simulations as the bare potential before adding the random potential fluctuations is the cut-out marked by the red square framing the range $x=0-1000nm$ and $y=0-1000nm$, compare also with (a). 
}

\end{figure*}

Fig.\ref{fig_pot}a shows the bare potential as used for the HF simulations in this paper. The intention is to create an arbitrarily shaped smooth bare edge potential that is neither linear nor parabolic. All simulations are mapped on a $199 \times 199$ grid. The indented direction of current flow is the x-direction as demonstrated further below. The left side of the slope resembles the smooth edge potential that is generated by the tail of a Gaussian peak, which's center lies outside of the active window as shown in Fig.\ref{fig_pot}c. The peak has a height of $V_0 = 600mV$, a width of $b_1=1000nm$ and is placed at $y_1 = -1000nm$ and in the centre of the x-axis at $x_0 = 500nm$. The system is terminated on the right by a considerably steeper slope. In particular, that slope is composed by the tails of 6 narrow Gaussian peaks of same height $V_0 = 600mV$, but a width of $b_2=100nm$ that get placed along the x-axis on the right side, also outside the active window at $y_2 = 1100nm$. Those peaks are placed at positions $x_1 = 0 nm$, $x_2 = 200 nm$, $x_3 = 400 nm$, $x_4 = 600 nm$, $x_5 = 800 nm$, $x_6 = 1000 nm$. Finally the random potential is added on top of it and it consists off a random superposition of $1000$ Gaussian peaks of random height $V_I$ between $-2mV$ and $+2 mV$ and of lateral range $b_I=30nm$. That finally results in overall potential fluctuations of about $\pm5mV$. 
\section{Results}

\begin{figure}[thb]
(a)\hspace*{0ex}
\includegraphics[angle=0,keepaspectratio=true,width=0.95\columnwidth,clip=true,trim=00 0 100 0]{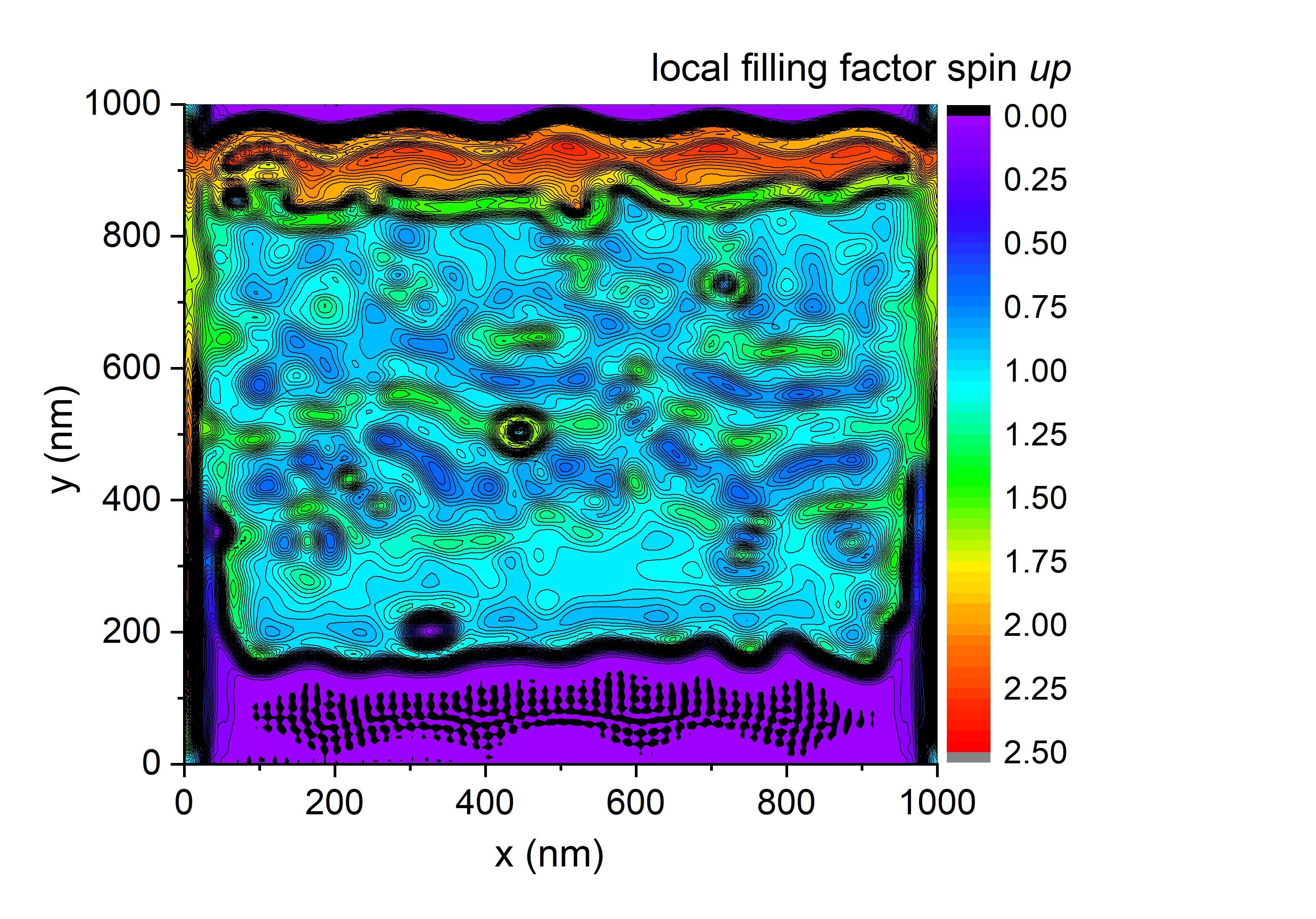}\\
%(b)\includegraphics[width=0.95\columnwidth,clip=true]{fitfreqSqrtBplot.jpg}
(b)\includegraphics[width=0.95\columnwidth,clip=true,trim=00 0 100 0]{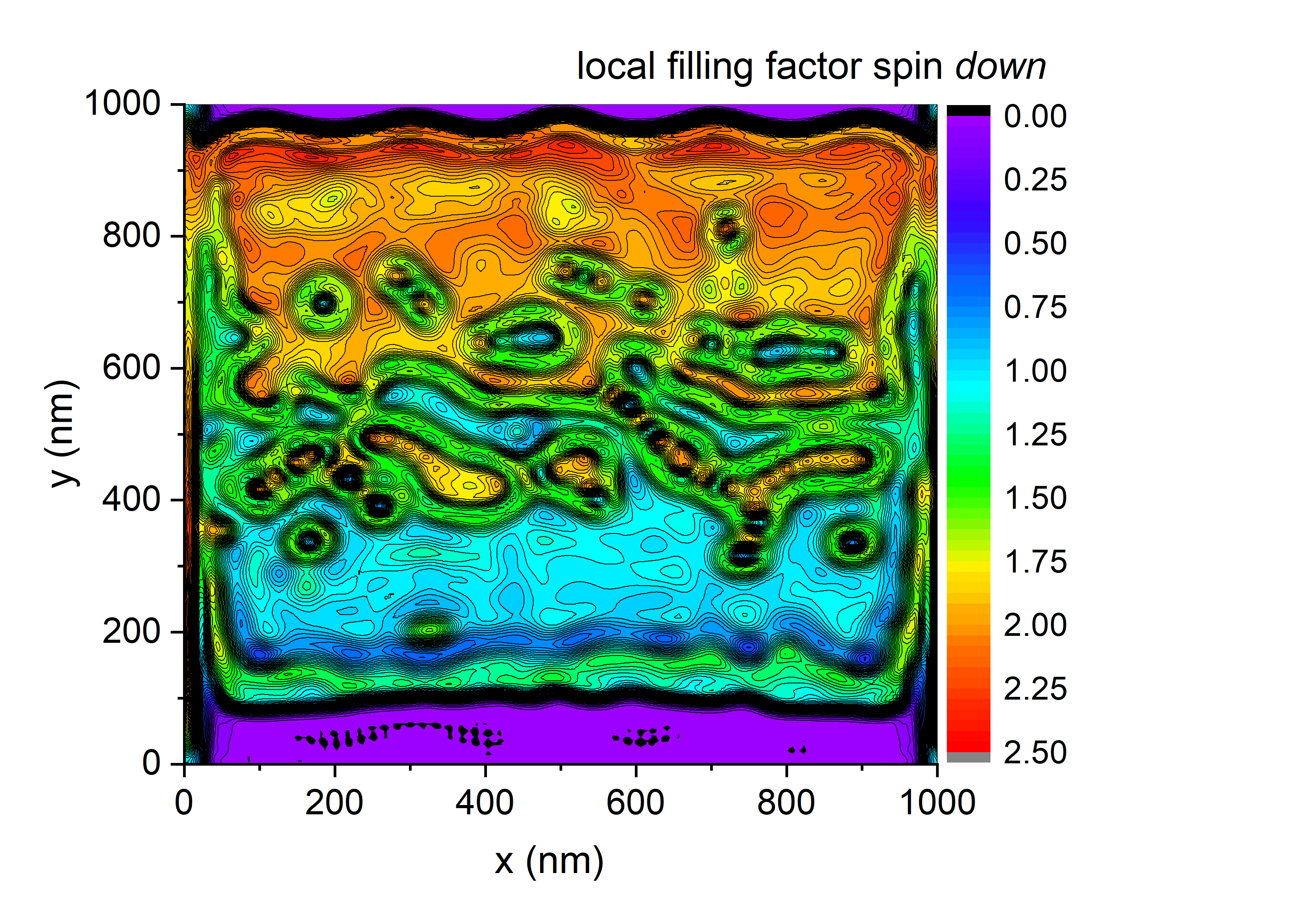}
% (b)\hspace*{-2ex}
% \includegraphics[width=0.33\columnwidth,clip=true,trim=20 10 10 20]{spec00075.png}
% (c)\hspace*{-2ex}
%\includegraphics[width=0.55\columnwidth,clip=true]{NMR2cropped.png} \\
\caption{\label{density}\
Lateral carrier density mapped on the filling factor scale $\nu_\uparrow$ for spin$\uparrow$ (a) and $\nu_\downarrow$ for spin$\downarrow$ (b) at $n=1.4 \cdot 10^{11}cm^{-2}$ and $B=2.5T$.
}
\end{figure}

Fig.\ref{fig_pot}b shows the screened potential that is calculated by the sum of the bare potential of Fig.\ref{fig_pot}a and the Hartree potential of the carrier distribution that in turn is obtained from the HF solution. As demonstrated in Fig.\ref{fig_pot}b, the wide soft edge potential gets completely flattened by screening, while the potential fluctuations remain almost non-screened as can be qualitatively seen by comparing Fig.\ref{fig_pot}a and Fig.\ref{fig_pot}b. For understanding this behavior, the lateral carrier density distribution $\nu_\uparrow$ for the spin$\uparrow$ LLs is shown in Fig.\ref{density}a and $\nu_\downarrow$ for the spin$\downarrow$  LLs is shown in Fig.\ref{density}b, while the total filling factor $\nu$ is given by $\nu = \nu_\uparrow + \nu_\downarrow$. For spin$\uparrow$ we get a dominating region of almost constant filling factor of $\nu_\uparrow =1$, except for a narrow region at the transition from the soft edge potential to the steep edge potential near the upper boundary, where a layer of filling factor $\nu_\uparrow=2$ starts to build up. This is also the region, where in macroscopic samples the bulk region would join in, but which has been omitted for this calculation because of size limitation. Near the lower edge an almost abrupt change from filling factor $\nu_\uparrow=1$ to $\nu_\uparrow=0$ appears. The spin$\downarrow$ LL is lower in energy and contains more electrons than the spin$\uparrow$ LL. However, the classically expected continuous increase of the carrier density in the spin$\downarrow$ LL in Fig.\ref{density}b is replaced by a widened region of also nearly constant integer filling $\nu_\downarrow=2$ as compared to the region of $\nu_\uparrow=2$ in Fig.\ref{density}a. Instead of a smooth decrease of the carrier density while approaching the edge (see appendix B), the $\nu_\downarrow=2$ layer breaks up into clusters of $\nu_\downarrow=2$ on top of a $\nu_\downarrow=1$ background that shrink in size and finally vanish while approaching the edge. Finally a homogeneous filling of $\nu_\downarrow=1$ is achieved that makes an abrupt change to $\nu_\downarrow=0$ near the lower edge. The classically expected smooth decrease of the carrier density while moving from the bulk towards the edge is replaced by a decreasing cluster density of constant filling $\nu_\downarrow=2$. In this way the classically expected smooth decrease of the carrier density exists only on average over long distances. For the screening of almost macroscopic long range potential modulations like the soft edge potential, this cluster structure is still sufficient to maintain a Thomas-Fermi like screening mechanism as expected also from semi-classical models like that one of CSG. However, the potential fluctuations of approximately same size or even smaller than the typical cluster size cannot be effectively screened anymore. Therefore the potential fluctuations are preserved like also assumed by the early models that neglect e-e interaction completely, while they have been using these potential fluctuations to explain the creation of a channel network. Pioneering work in this context has been done by Chalker and Coddington\cite{ChaC88} and their so called Chalker-Coddington-network has been frequently used also by others\cite{RN433, Kramer2005}. However, as explained earlier\cite{Romer2021TheRegime,OswaldPRB2017, OswR17}, in our case the channels consist of the cluster boundaries that are visible as green colored thin stripes in Fig.\ref{density}b. Hence, the created channel network follows not exactly, but roughly the equipotentials of the potential fluctuations. However, that does not necessarily harm their overall statistical character as assumed in various network models in context with the scaling theory\cite{Huckestein1995ScalingEffect} and therefore the scaling results obtained by the early non-interacting models in the past will remain valid. The compressibility in our case becomes restricted to a growing or shrinking of the clusters by adding electrons to or removing electrons just from those narrow green colored cluster boundaries in Fig.\ref{density}b that get shaped by the details of the Landau basis functions\cite{HasCFS12, OswR17}. Like assumed already by the early non-interacting single particle models, also in our case the width of the channels remains of the order of the cyclotron radius $R_c =\sqrt{(2j+1)\hbar /e B}$, with LL-index $j=0,1,2,...$ which is a quantity that has been found to be important also for the stripe and bubble regime\cite{Goerbig2004a, Oswald2020}.

%%%%%%%%%%%%%%%%%%%%%%%%%%%%%%%%%%%%%%%%%%%%%%%%%%%%%%%%%%%%%%%%

\begin{figure}[thb]
(a)\hspace*{0ex}
\includegraphics[angle=0,keepaspectratio=true,width=0.95\columnwidth,clip=true,trim=20 0 -00 0]{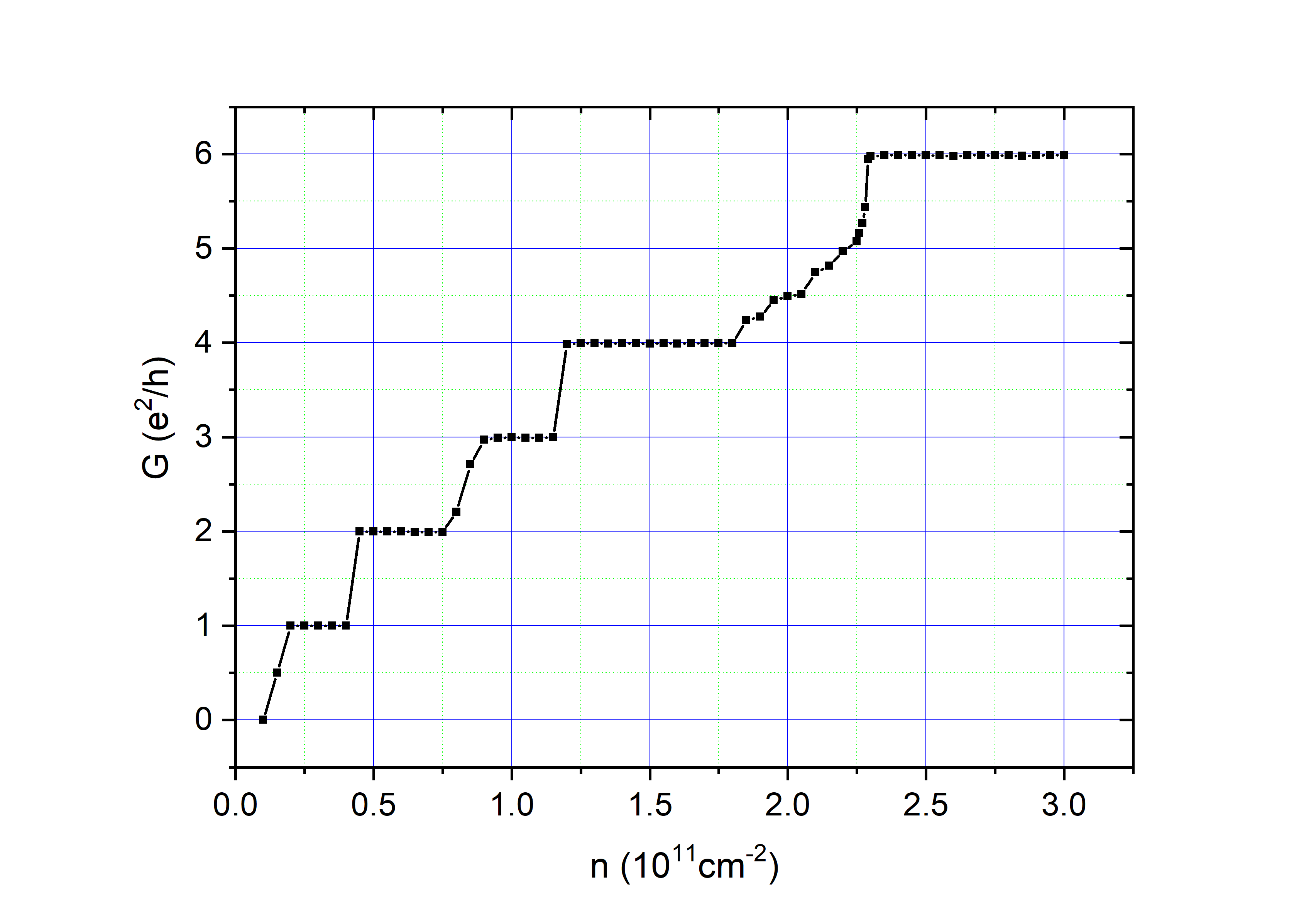}\\
%(b)\includegraphics[width=0.95\columnwidth,clip=true]{fitfreqSqrtBplot.jpg}
(b)\includegraphics[width=0.95\columnwidth,clip=true,trim=00 0 100 0]{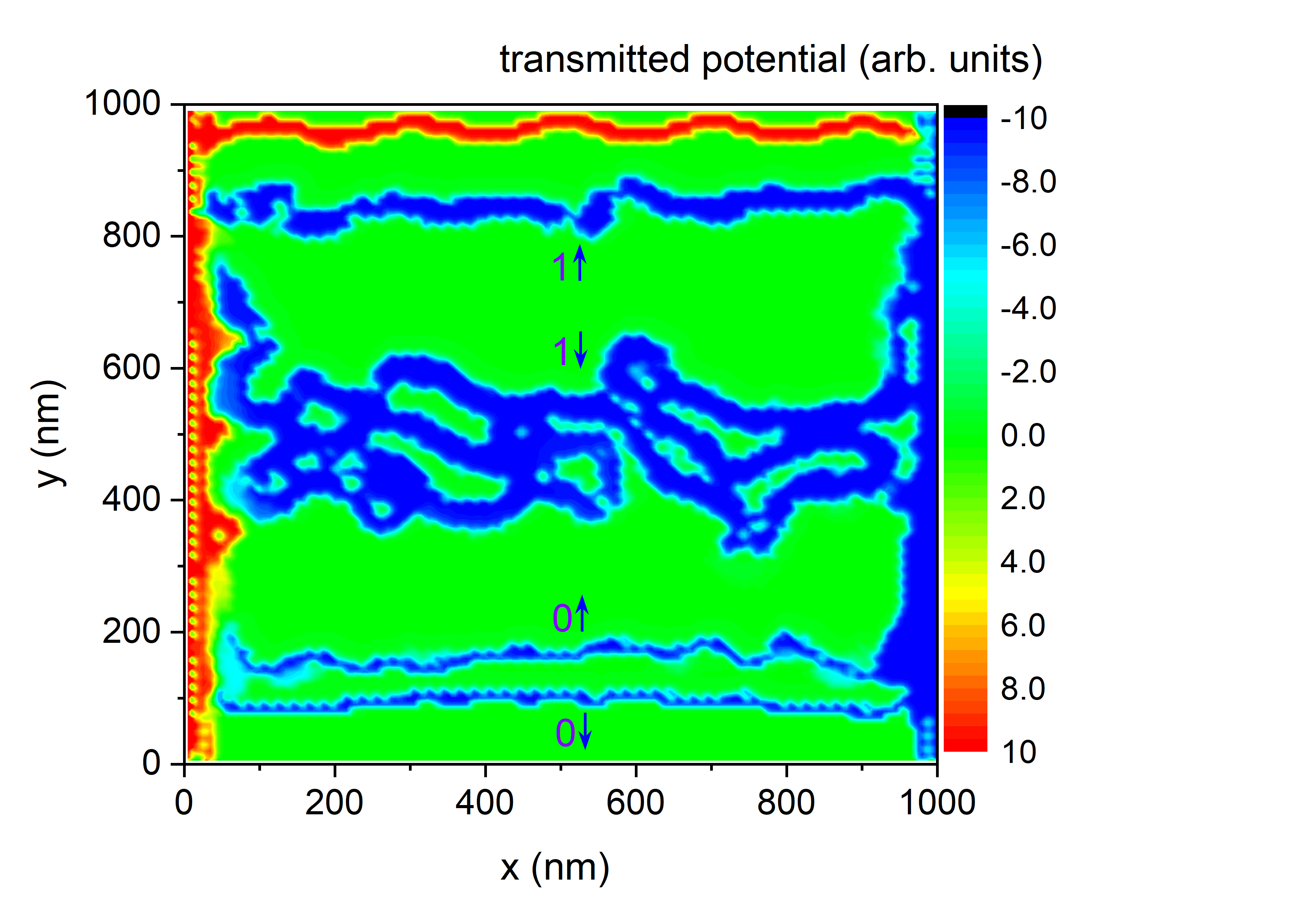}
% (b)\hspace*{-2ex}
% \includegraphics[width=0.33\columnwidth,clip=true,trim=20 10 10 20]{spec00075.png}
% (c)\hspace*{-2ex}
%\includegraphics[width=0.55\columnwidth,clip=true]{NMR2cropped.png} \\
\caption{\label{transport}
(a) Two-point conductance versus average carrier density at constant magnetic field $B=2.5T$ (b) Lateral distribution of the transmitted non-equilibrium potential as obtained from the NNM at $n=1.4 \cdot 10^{11}cm^{-2}$ and $B=2.5T$. The labels 0$\uparrow$, 0$\downarrow$ and 1$\uparrow$, 1$\downarrow$ denote the channels generated by the lowest spin-up and spin-down LLs as well as the second spin-up and spin-down LLs respectively.}
\end{figure}

We have also addressed non-equilibrium transport using the non-equilibrium network model (NNM) as introduced earlier\cite{Oswald2006} and explained repeatedly\cite{Osw16,OswaldPRB2017}. Fig.\ref{transport}a shows the two-point conductance in x-direction versus average carrier density at constant magnetic field of $B=2.5T$. One can clearly see that there appear spin-resolved conductance steps. Tentatively, the odd plateaus appear less pronounced than the even plateaus and the $\nu=5$ plateau is even missing. The main reason for the missing $\nu=5$ plateau is the vanishing bulk region that does not provide a sufficient size for a fully-fledged exchange enhanced spin splitting and thus, we obtain not sufficiently well separated edge channels for the two spin states within the same LL. This gets more critical with increasing cyclotron radius $R_c$ at higher LLs.  Fig.\ref{transport}b shows the lateral distribution of the transmitted excitation potentials obtained by the NNM. Note, the transmitting channels for the excitation potentials must not be confused with the current path, which remains undefined within the regions between channels transmitting different potentials\cite{Oswald2006, Osw16}. The cirality is clock-wise, which means the high potential (introduced along the y-axis on the left and colored in red) gets transmitted along the upper edge from left to right. The low potential (introduced along the y-axis on the right side and colored in blue) gets transmitted from right to left along the lower edge, which constitutes the soft edge that takes up almost the whole sample area below the upper boundary in this simulation. In the region of the soft edge we see indeed 4 well separated channels in blue color, while there appears only one red channel to be visible near the upper edge. This is because at the upper boundary we have a steep hard wall like edge potential that cannot be screened and therefore all 4 red channels cannot be sufficiently separated and they appear on top of each other. In Fig.\ref{fig_pot}b there is an almost macroscopically wide flat region. That region consists of the flattened edge potential terrace, which merges with the vanishing narrow bulk region, but they cannot be distinguished from each other on this scale. However, the uppermost blue channel in Fig.\ref{transport}b can be identified as the boundary between the vanishing bulk region and the adjacent edge potential terrace that is created by the screening of the wide soft edge potential. Looking at Figs.\ref{fig_pot}b, \ref{density}b and \ref{transport}b together, we see that the spin$\downarrow$ LL creates a wide channel network in the middle of the potential terrace(Fig.\ref{transport}b). Looking at Fig.\ref{density}b we see the $\nu_\downarrow = 2$ clusters with their boundaries (green color) that correspond with the blue colored channel network in Fig.\ref{transport}b. Towards the lower edge there exist two more narrow blue channels that result from the $\nu_\uparrow = 1$ and $\nu_\downarrow = 1$ cluster boundaries that complete the system of four channels as required for the fourth conductance plateau at $n=1.4 \cdot 10^{11}cm^{-2}$. 

\section{Discussion}

All together, our results have strong similarities with the early models of the IQHE based on non-interacting electrons in narrow channels and channel networks. However, also the main features of wide compressible stripes such as the flattening of macroscopic soft edge potentials to an edge potential terrace is exactly represented. But most importantly, in contrast to the theory of CSG\cite{Chklovskii1992}, our model preserves the existence of the narrow channels that generate a random network of quantum channels of top of those wide stripes. Oppositely to CSG (see also Appendix D), the narrow channels from the early non-interacting models do not widen up into wide compressible stripes by taking into account e-e interaction. Instead, the compressible stripes become a mixture of clusters of full and empty LLs while their boundaries create a network of still narrow quantum channels. This is a very fundamental result that decisively rests on many-body interactions (see also Appendix B,C). At the same time this behaviour embodies a hallmark of many body physics: Low energy excitations close to the many-body ground state tend to show up as non-interacting quasi particles, while the many-body interactions are absorbed by the many particle ground state. Such low energy excitations are only possible at the boundaries of the clusters that finally act as quantum channels. Pascher et al performed scanning gate investigations of quantum point contacts in the QH-regime\cite{PasRIE14} and their analysis strongly supports our results: The shape and width of the locally initiated conductance steps fit more closely to a model based on non-interacting electrons in narrow channels than to models based on e-e interaction according to CSG. Our results are also strongly supported by the fact that a Tomonaga–Luttinger-liquid (TLL) behaviour of edge excitations is observed\cite{Hashisaka2018TomonagaLuttinger-liquidChannels, Rodriguez2020RelaxationLiquid}. In this context our results provide the necessary basis for the applicability of the TLL theory for QH edge excitations. The bosonic character of the low-energy excitations, as one major ingredient of the TLL theory, has been proposed also independently in context with our non-equilibrium network model\cite{RN38, Osw16}. On the basis of an intuitive understanding in terms of an effective non-interacting single particle picture, the electrons in a cluster "see" the cluster boundaries as an effective hard wall potential (see Appendix A). In this way our results, obtained about 40 years later, finally provide a physically realistic basis for the early edge state approach based on hard wall potentials by Halperin\cite{Hal82} and B\"uttiker\cite{But88}. Other early non-interacting single particle models based on a phenomenological mixture of two quantum Hall liquid phases, like formulated by Dykhne and Ruzin\cite{Dykhne1994TheoryModel}, or Ruzin and Feng\cite{Ruzin1995} also get a much more realistic theoretical basis by our results that introduce a similar mixture of full and empty clusters. The latter has already been discussed in context with our NNM\cite{Oswald2006}. Another prominent phenomenon that supports the correctness of our results is the observation of high quality quantum interference oscillations of edge modes that would be hardly possible without the formation of narrow edge modes like discussed in Appendix E. Last but not least, also the already mentioned success of the scaling theory\cite{Huckestein1995ScalingEffect} justifies the existence of narrow channels in the IQHE regime.       

In conclusion, we find that our results bring the IQHE and the fractional quantum Hall effect (FQHE) much closer to each other than it is commonly understood so far, although we do not yet directly address the FQHE. While for the FQHE the low-energy excitations of the many particle ground state show up as non-interacting quasi particles that are known as composite Fermions\cite{Jain1995}, for the IQHE the electrons seem to appear as their own non-interacting quasi particles resulting from low energy excitations of the many particle ground state. Note, in this context it is well known that the FQHE can also be interpreted as an IQHE of the composite Fermions in an effective magnetic field\cite{Jain1995, Jai15}. Also the re-entrant IQHE that appears as a switching back and forth between the IQHE and the FQHE in ultra clean QH-systems\cite{Eisenstein2002,Deng2012c} indicates the strong connection between the IQHE and the FQHE. All together, our results show that a significant part of the new physics that rests on many-body interactions and has been attributed mainly to the discovery of the FQHE, is already driving also the IQHE.

\section{Acknowledgments}

\begin{figure}[tb]
%\mbox{ }  \hfill Hartree \hfill non-interacting \hfill  \mbox{ }\\
%\hspace*{-0.8\columnwidth} $\uparrow$ \hspace{0.8\columnwidth}$\downarrow$
\includegraphics[width=0.45\textwidth,clip=true,trim=0 00 0 0]{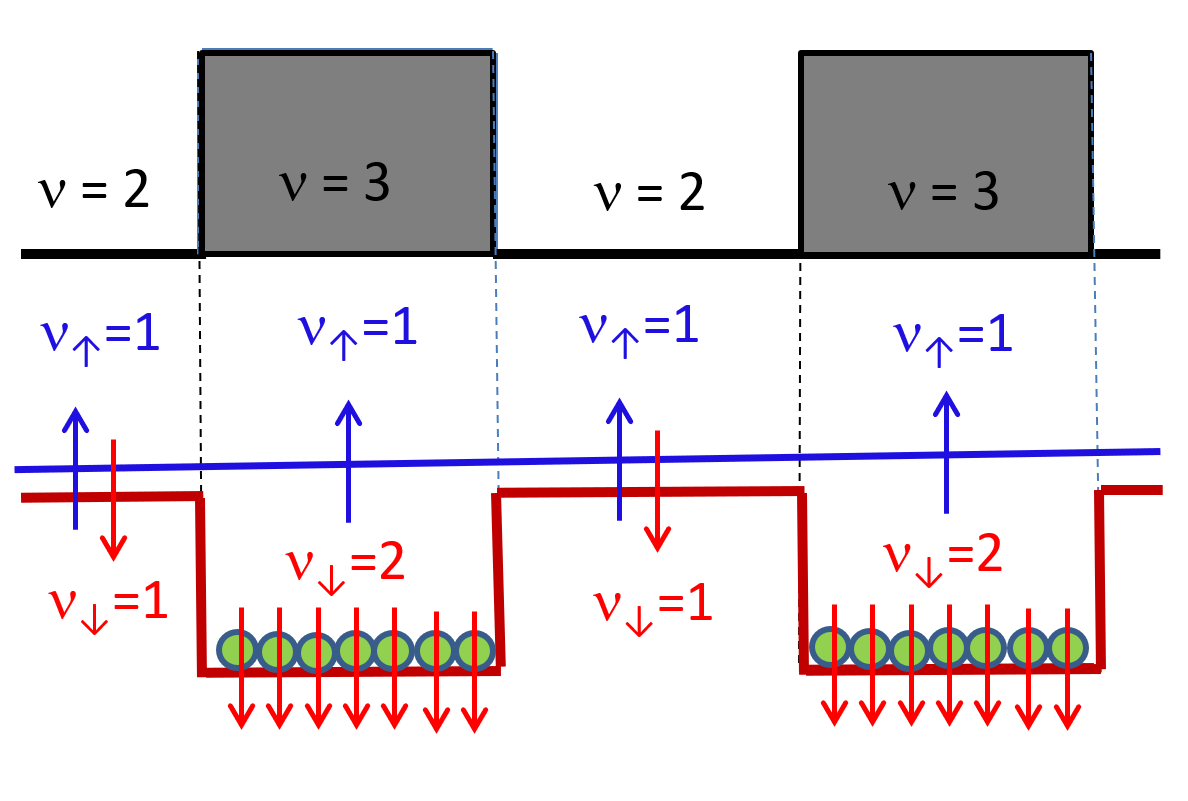}
%(b) \includegraphics[width=0.45\textwidth,clip=true,trim=140 00 140 00]{EffectivePot.png}\\
%(c) \includegraphics[width=0.65\textwidth,clip=true,trim=000 70 00 0]{generate_pot.png}
%(d) $\downarrow$\includegraphics[width=0.45\textwidth]{CD_nu_450S1-00_.png}
\caption{
\label{EffectivePot}
Schematic representation of the many-body interactions based on an intuitive effective single electron picture for cluster formation (see also text). The upper part schematically shows the lateral variation of the filling factor while crossing e.g. $\nu = 3$ clusters. The lower part shows schematically the associated lateral change of the exchange enhanced spin splitting with blue for the spin-up and red for the spin-down levels. The enhanced spin splitting within the odd $\nu = 3$ clusters creates effective potential wells that keep the electrons within the clusters.
}
\end{figure}

J.O. thanks Karl Flicker for valuable technical sup-
port for the MUL high performance cluster in Leoben, where the HF calculations have been performed.
\newline
{\bf APPENDIX A: Intuitive non-interacting single particle picture}
\newline
\newline
Fig.\ref{EffectivePot} is a schematic illustration, how the effect of the exchange interaction can be easily understood on the basis of an intuitive effective non-interacting single particle picture. As already mentioned in the paper, for the assumed non-interacting single particles we have to consider an effective potential that consists of the sum of the Hartree potential and the effect of the lateral variation of the exchange enhanced spin-splitting that in turn depends on the lateral variation of the total filling factor $\nu$. In the upper part of Fig.\ref{EffectivePot} there is shown the simplified schematic lateral variation of the total filling factor if crossing $\nu_\downarrow=2$ clusters ($\nu=3$) like in Fig.2b. In the lower part of the figure there is drawn the corresponding relative variation of the spin$\uparrow$ (blue) and the spin$\downarrow$ (red) levels. At even filling factors the spin levels are close to each other because of low g-factor enhancement and at odd filling factors the spin levels get pushed apart due to a large exchange enhanced spin splitting. This creates a positive feedback loop for the self consistent carrier redistribution. Hence, we get jumps in the spin levels that create effective potential wells which self consistently stabilize the clusters even against the Hartree repulsion. Simply speaking, the boundaries of the clusters act like hard wall potentials keeping the effective (quasi) single electrons in the cluster. Like shown in the paper, the boundaries of those clusters create narrow channels for low energy excitations in transport. This mechanism looks striking similar to the creation of edge channels in the presence of a hard wall potential like proposed by Halperin\cite{Hal82}.   
\newline
\newline
{\bf APPENDIX B: Hartree results}
\newline
\begin{figure}[tb]
%\mbox{ }  \hfill Hartree \hfill non-interacting \hfill  \mbox{ }\\
%\hspace*{-0.8\columnwidth} $\uparrow$ \hspace{0.8\columnwidth}$\downarrow$
\includegraphics[width=0.43\textwidth,clip=true,trim=60 30 120 00]{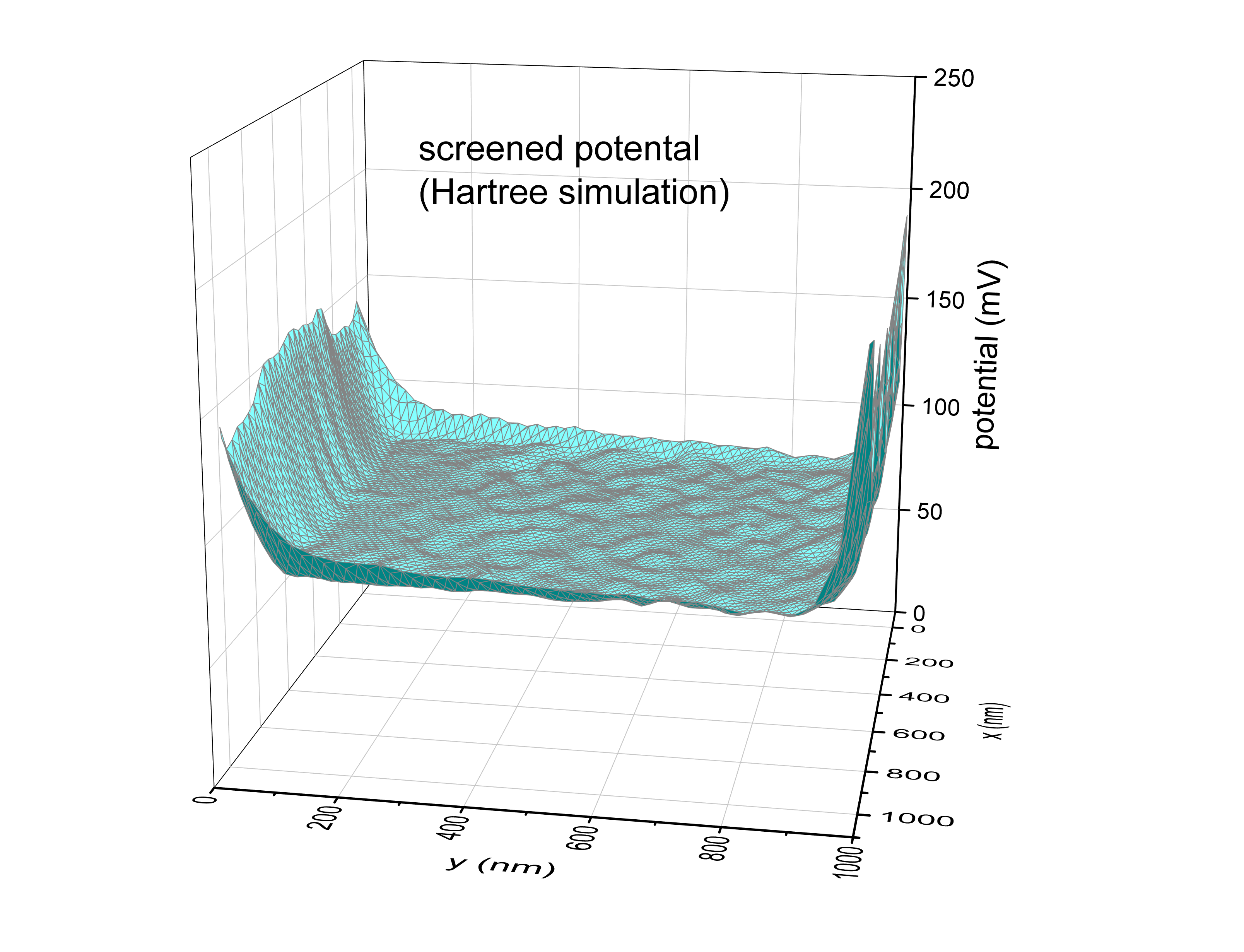}\\
%(c) \includegraphics[width=0.65\textwidth,clip=true,trim=000 70 00 0]{generate_pot.png}
%(d) $\downarrow$\includegraphics[width=0.45\textwidth]{CD_nu_450S1-00_.png}
\caption{
\label{S_fig_pot}
Screened potential as obtained from the pure Hartree solution for exactly the same parameters as used in the paper for Fig.1b. 
}
\end{figure}
\begin{figure}[tb]
%\mbox{ }  \hfill Hartree \hfill non-interacting \hfill  \mbox{ }\\
%\hspace*{-0.8\columnwidth} $\uparrow$ \hspace{0.8\columnwidth}$\downarrow$
\includegraphics[width=0.45\textwidth,clip=true,trim=20 00 120 0]{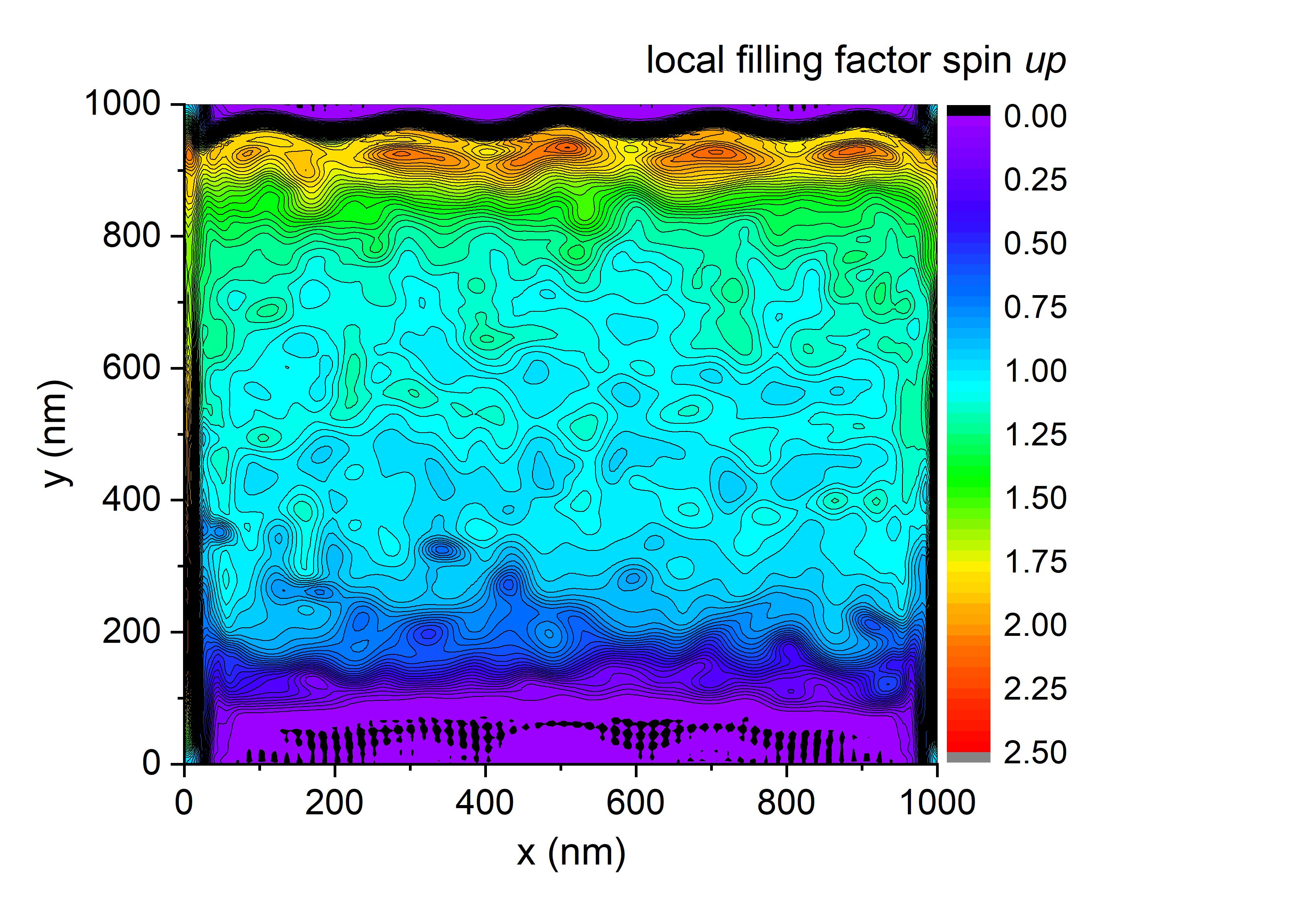}
%(c) \includegraphics[width=0.65\textwidth,clip=true,trim=000 70 00 0]{generate_pot.png}
%(d) $\downarrow$\includegraphics[width=0.45\textwidth]{CD_nu_450S1-00_.png}
\caption{
\label{S1_density}
Lateral carrier density based on a pure Hartree calculation mapped on the filling factor scale for spin$\uparrow$ at $n=1.4 \cdot 10^{11}cm^{-2}$ and $B=2.5T$. 
}
\end{figure}
\begin{figure}[tb]
%\mbox{ }  \hfill Hartree \hfill non-interacting \hfill  \mbox{ }\\
%\hspace*{-0.8\columnwidth} $\uparrow$ \hspace{0.8\columnwidth}$\downarrow$
\includegraphics[width=0.45\textwidth,clip=true,trim=20 00 120 00]{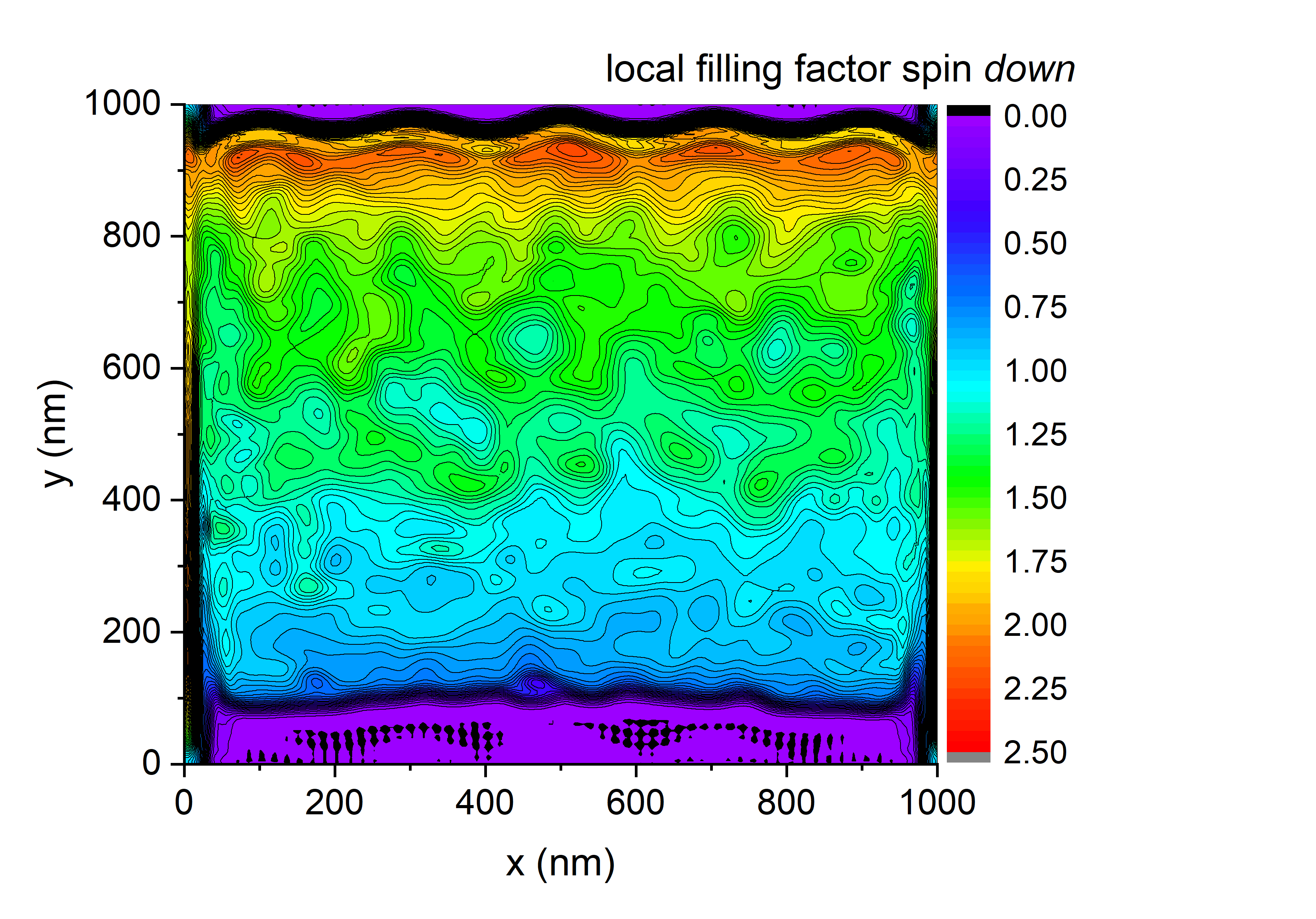}\\
%(c) \includegraphics[width=0.65\textwidth,clip=true,trim=000 70 00 0]{generate_pot.png}
%(d) $\downarrow$\includegraphics[width=0.45\textwidth]{CD_nu_450S1-00_.png}
\caption{
\label{S2_density}
Lateral carrier density based on a pure Hartree calculation mapped on the filling factor scale for spin$\downarrow$ at $n=1.4 \cdot 10^{11}cm^{-2}$ and $B=2.5T$. 
}
\end{figure}

Fig.\ref{S_fig_pot} shows the screened potential based on pure Hartree interaction that turns out to be almost identical to that one based on the HF results (compare Fig.\ref{fig_pot}b). Again, the soft edge potential slope gets completely fattened. However, also the potential fluctuations get hardly screened as compared to the HF results. While the absence of screening of the mesoscopic fluctuations in the HF solution is a consequence of the typical cluster size that does not allow a sufficient screening of details with smaller size, such screening could in principle be expected for the Hartree case. However, also for the Hartree case there exists a "resolution limit" or in other words, a "high frequency cut-off" for screening and that typical length scale is connected to the lateral extension of the single particle wave functions. That is of the order of cyclotron radius $R_c$, which for our calculations appears to be $R_c \approx 30nm $. This is about the same lateral extension like the impurity potentials used to create the random potential. Therefore we expect that the screening of the potential fluctuations on the basis of the Hartree model would become more pronounced only for higher magnetic fields, while the screening of the fluctuations on the basis of the HF model will remain suppressed. Anyway, the major contrast between HF and Hartree results turns up in the lateral carrier density distribution, which for the Hartree case is shown in Fig.\ref{S1_density} for spin$\uparrow$ and Fig.\ref{S2_density} for spin$\downarrow$. In the particular case mainly the spin$\downarrow$ electrons are creating the wide flat potential terrace, while at the same time for the HF interactions they create also a network of narrow channels on top of it (compare Fig.\ref{density}b). In contrast, using just Hartree interaction, as shown in Fig.\ref{S2_density}, these electrons create just a more or less smooth carrier density variation while approaching the edge. The color-coding enhances the filling factor range around half-filling by the color green and it is easily seen that in Fig.\ref{S2_density} this region extends over at least half of the area indicating an almost macroscopic wide compressible region. That is exactly what also the CSG model delivers: Instead of getting jumps in the density, we get a smooth variation while the screened edge potential resembles a flat terrace. In contrast, for Fig.\ref{density}b this compressible (green) region is replaced by a network of narrow channels. Hence, the results of CSG are clearly reproduced by the Hartree calculations, but the narrow channels of the HF solution have disappeared. As a consequence, also the theoretical basis for applying e.g. the scaling theory in terms of a random network of narrow channels, as well as the theoretical basis for the also quite successful TLL theory is gone. Obviously only many-particle physics preserves the theoretical basis for applying those most successful theories. 
\newline
\newline
{\bf APPENDIX C: Further Hartree-Fock results}
\newline
\begin{figure}[tb]
%\mbox{ }  \hfill Hartree \hfill non-interacting \hfill  \mbox{ }\\
%\hspace*{-0.8\columnwidth} $\uparrow$ \hspace{0.8\columnwidth}$\downarrow$
\includegraphics[width=0.45\textwidth,clip=true,trim=20 00 120 0]{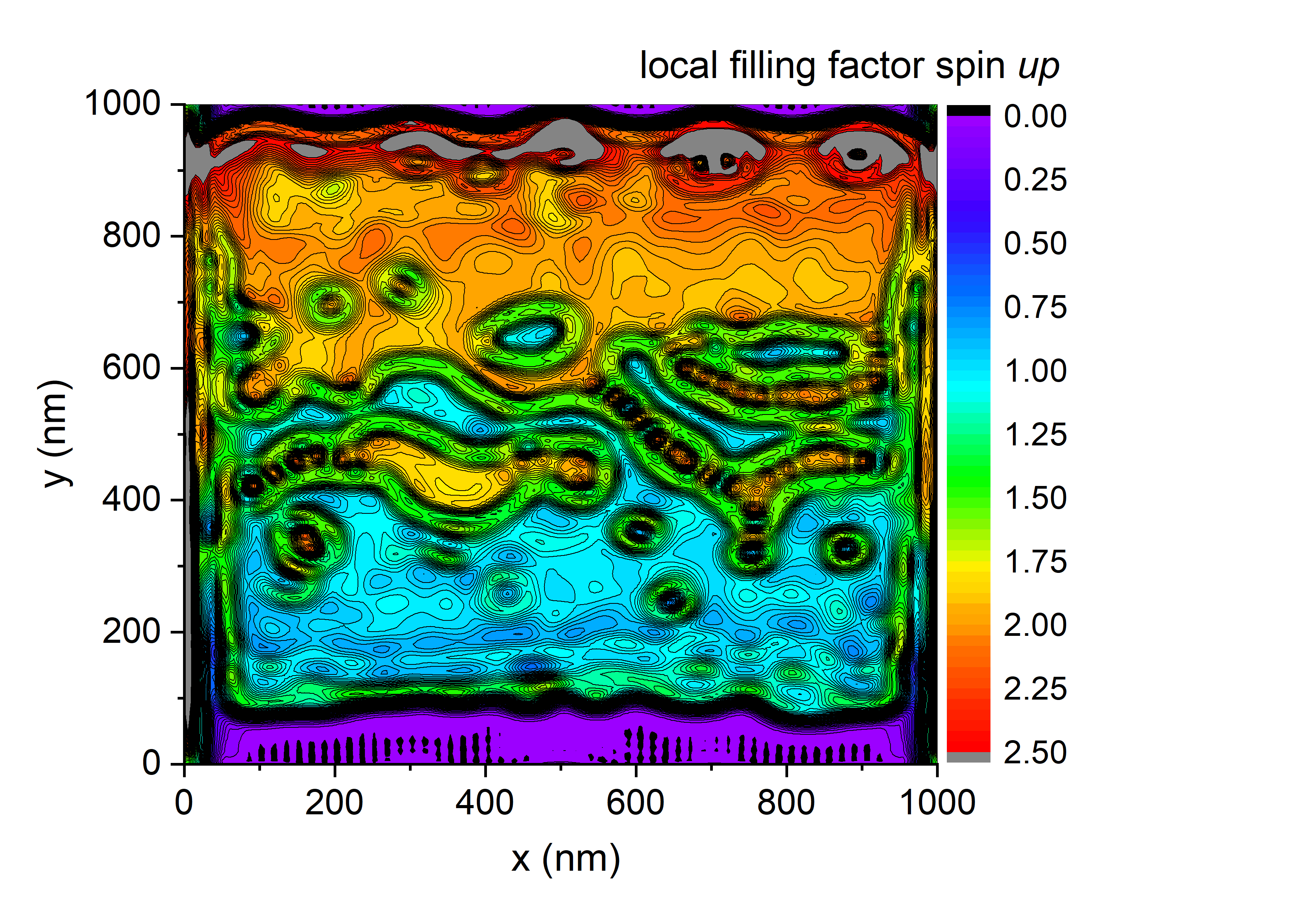}
%(c) \includegraphics[width=0.65\textwidth,clip=true,trim=000 70 00 0]{generate_pot.png}
%(d) $\downarrow$\includegraphics[width=0.45\textwidth]{CD_nu_450S1-00_.png}
\caption{
\label{S1_density_nu5}
Lateral carrier density mapped on the filling factor scale for spin $\uparrow$ at $n=2.0 \cdot 10^{11}cm^{-2}$ and $B=2.5T$. 
}
\end{figure}

\begin{figure}[tb]
%\mbox{ }  \hfill Hartree \hfill non-interacting \hfill  \mbox{ }\\
%\hspace*{-0.8\columnwidth} $\uparrow$ \hspace{0.8\columnwidth}$\downarrow$
\includegraphics[width=0.45\textwidth,clip=true,trim=20 00 120 00]{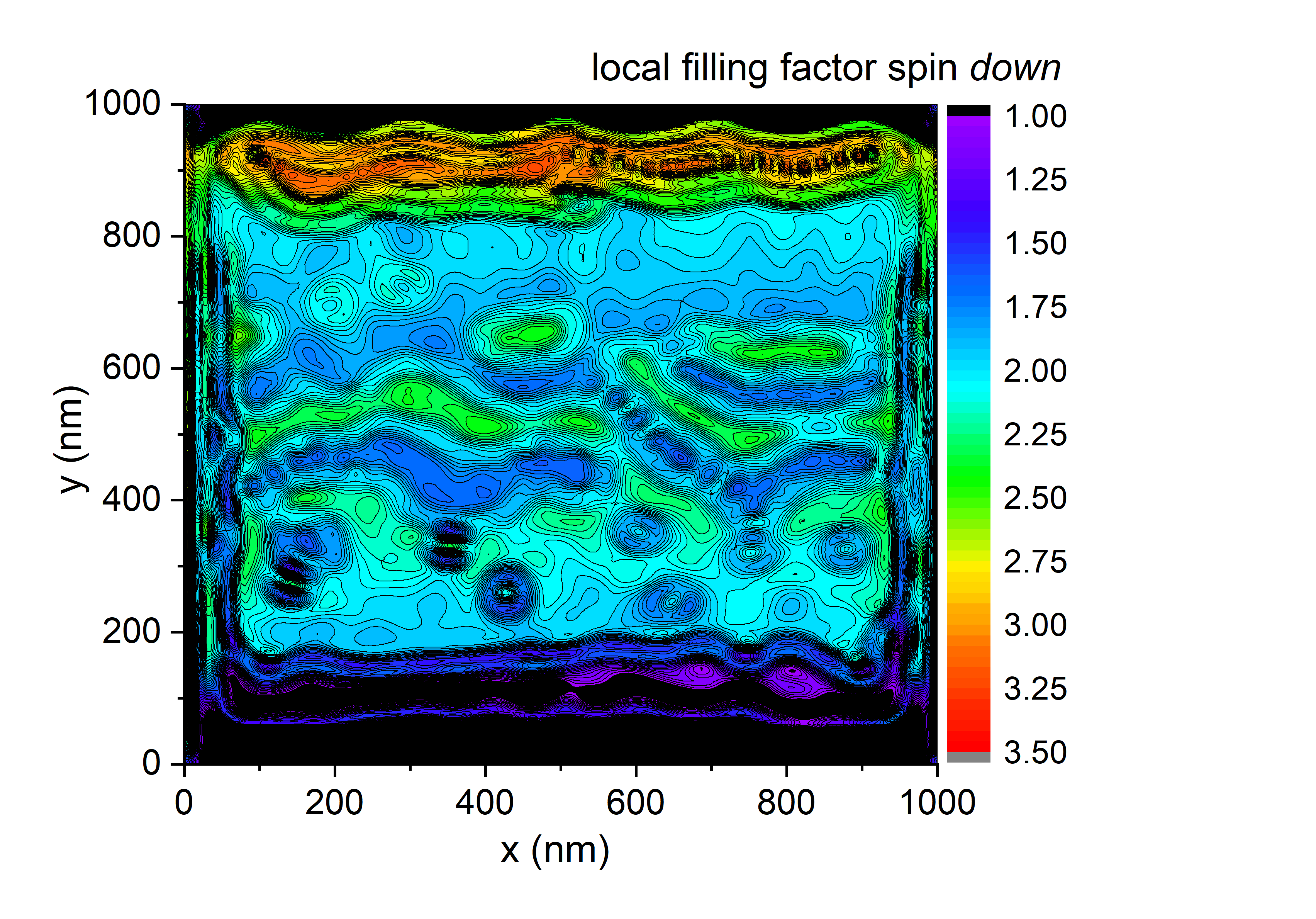}\\
%(c) \includegraphics[width=0.65\textwidth,clip=true,trim=000 70 00 0]{generate_pot.png}
%(d) $\downarrow$\includegraphics[width=0.45\textwidth]{CD_nu_450S1-00_.png}
\caption{
\label{S2_density_nu5}
Lateral carrier density mapped on the filling factor scale for spin$\downarrow$ at $n=2.0 \cdot 10^{11}cm^{-2}$ and $B=2.5T$. 
}
\end{figure}

\begin{figure}[tb]
%\mbox{ }  \hfill Hartree \hfill non-interacting \hfill  \mbox{ }\\
%\hspace*{-0.8\columnwidth} $\uparrow$ \hspace{0.8\columnwidth}$\downarrow$
\includegraphics[width=0.45\textwidth,clip=true,trim=20 00 120 0]{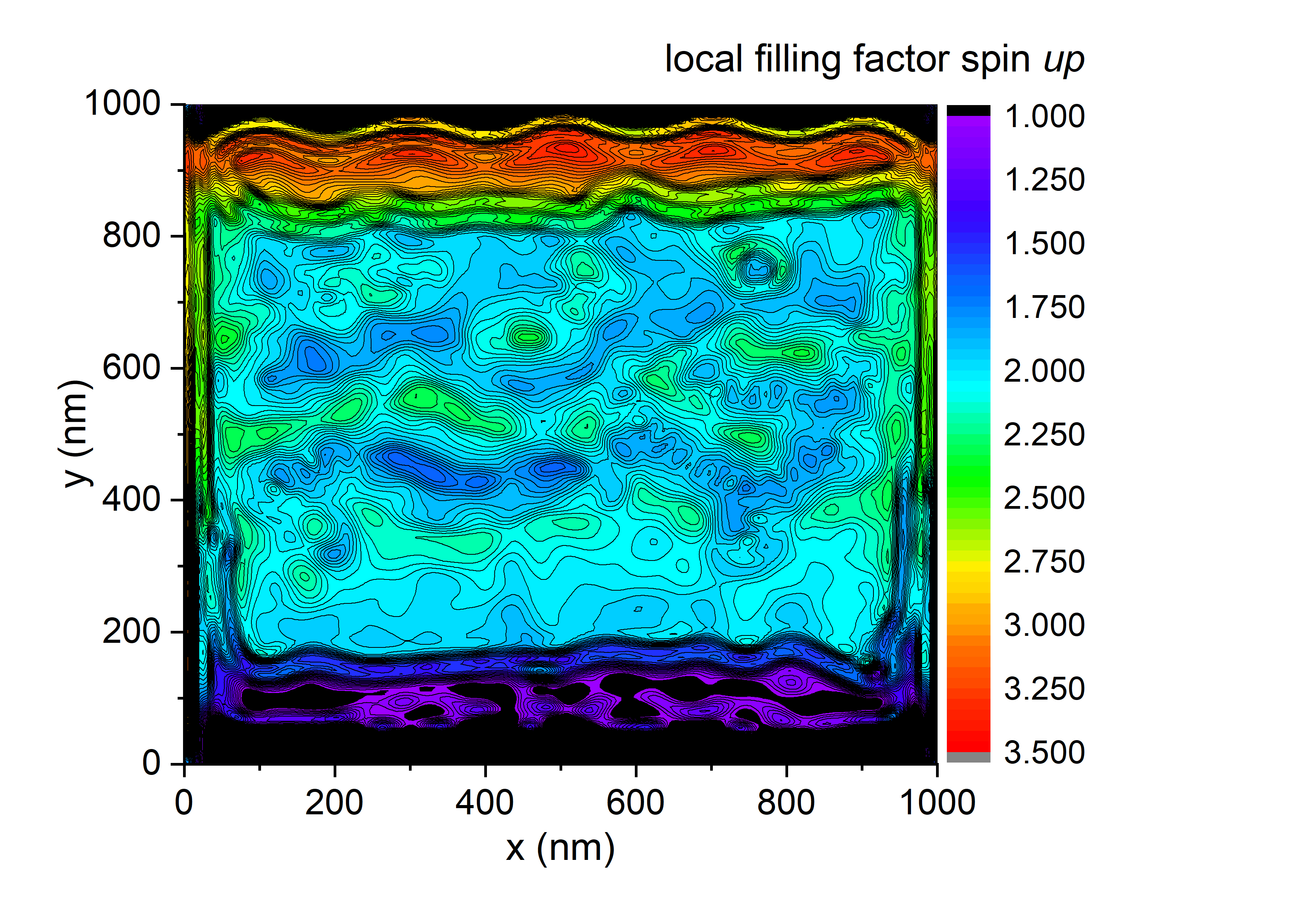}
%(c) \includegraphics[width=0.65\textwidth,clip=true,trim=000 70 00 0]{generate_pot.png}
%(d) $\downarrow$\includegraphics[width=0.45\textwidth]{CD_nu_450S1-00_.png}
\caption{
\label{S1_density_nu6}
Lateral carrier density mapped on the filling factor scale for spin$\uparrow$ at $n=2.6 \cdot 10^{11}cm^{-2}$ and $B=2.5T$. 
}
\end{figure}

\begin{figure}[tb]
%\mbox{ }  \hfill Hartree \hfill non-interacting \hfill  \mbox{ }\\
%\hspace*{-0.8\columnwidth} $\uparrow$ \hspace{0.8\columnwidth}$\downarrow$
\includegraphics[width=0.45\textwidth,clip=true,trim=20 00 120 00]{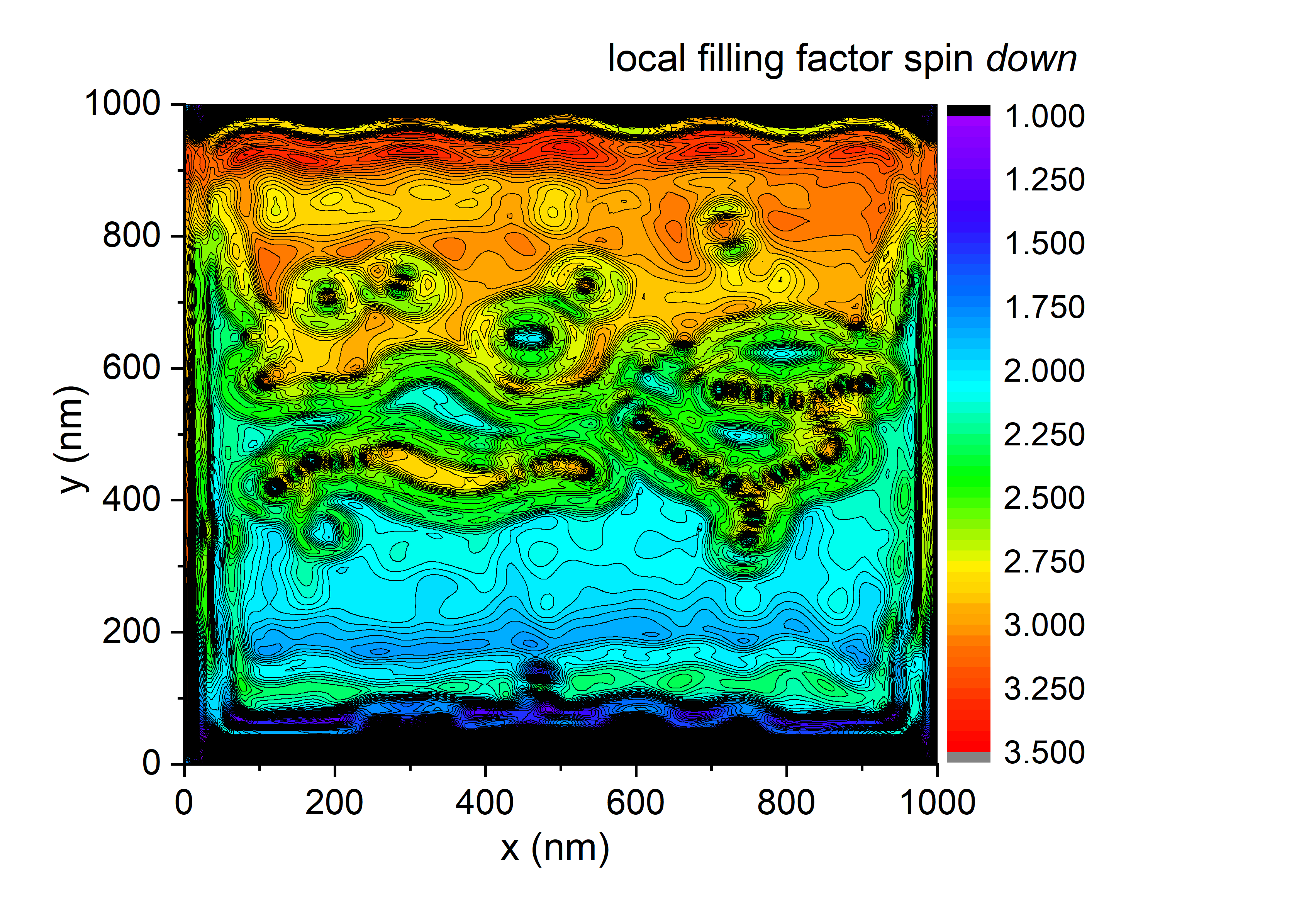}\\
%(c) \includegraphics[width=0.65\textwidth,clip=true,trim=000 70 00 0]{generate_pot.png}
%(d) $\downarrow$\includegraphics[width=0.45\textwidth]{CD_nu_450S1-00_.png}
\caption{
\label{S2_density_nu6}
Lateral carrier density mapped on the filling factor scale for spin$\downarrow$ at $n=2.6 \cdot 10^{11}cm^{-2}$ and $B=2.5T$. 
}
\end{figure}

\begin{figure*}[tb]
%\mbox{ }  \hfill Hartree \hfill non-interacting \hfill  \mbox{ }\\
%\hspace*{-0.8\columnwidth} $\uparrow$ \hspace{0.8\columnwidth}$\downarrow$
\includegraphics[width=0.80\textwidth,clip=true,trim=00 00 00 0]{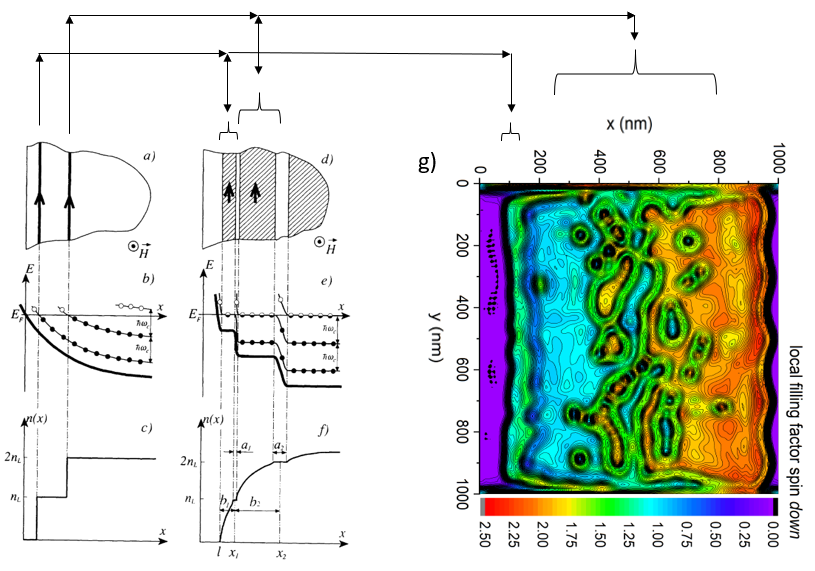}
%(c) \includegraphics[width=0.65\textwidth,clip=true,trim=000 70 00 0]{generate_pot.png}
%(d) $\downarrow$\includegraphics[width=0.45\textwidth]{CD_nu_450S1-00_.png}
\caption{
\label{CSGvsHF}
a-f are the original figure from the paper of CSG and g is Fig.2b of the actual paper with flipped x- and y - axis for better comparison. a-c shows schematically the situation as obtained from non-interacting electrons; d-f shows schematically the situation as obtained from single particle interaction according to CSG; g shows our HF results from the actual paper. The arrows on the top show how the two narrow edge channels on the left develop into associated stripes according to CSG or develop into a network of narrow channels (green color) due to many body interactions on the right. The bulk region in g is missing as explained in the paper. The associated soft edge is on the left for all a-g. 
}
\end{figure*}

 In this section we demonstrate that the cluster and network formation as shown for the $\nu_\downarrow=2$ level in Fig.\ref{density}b, works also for the $\nu_\uparrow=2$ level. For this purpose we use a higher average carrier density of $n=2.0\cdot 10^{11}cm^{-2}$ and the results are shown in Fig.\ref{S1_density_nu5} and Fig.\ref{S2_density_nu5}. In this case the role of spin$\uparrow$ and spin$\downarrow$ seems some how swapped as compared to Fig.2a and Fig.2b. Note, in order to keep the same color coding like in Fig.\ref{density}, the filling factor scale in Fig.\ref{S1_density_nu5} is kept the same like in Fig.\ref{density}b although the maximum density in Fig.\ref{S1_density_nu5} partly exceeds that scale (see some grey regions near the upper edge). The filling factor scale in Fig.\ref{S2_density_nu5} had to be shifted up and offset by one for the same reason. Finally, the channels in Fig.\ref{S1_density_nu5} arrange almost like in Fig.2b. Looking at the onset of the $\nu_\downarrow=3$ layer near the upper boundary of in Fig.\ref{S2_density_nu5}, we see that it has a more complex internal double structure as compared to Fig.\ref{density}a. That is due to the more complex Landau basis functions for the 3rd LL. It is also nicely seen that these channels appear not so well separated by the narrow bulk region and hence, the odd (spin-split) conductance plateau $\nu =5$ in Fig. 3a can hardly develop. 

Fig.\ref{S1_density_nu6} and Fig.\ref{S2_density_nu6} show a similar situation like in Fig.2a and Fig.2b, but in the 3rd LL at the $\nu = 6$ plateau of the transport data shown in Fig.3a. In this case the channel network is created again in the spin$\downarrow$ level, but for the 3rd LL the channels are wider and have a characteristic internal double mode structure due to the more complex single particle wave functions that have to be used for composing the many particle wave functions in the HF procedure. However, they arrange roughly in the same way like in Fig.2b. Note, in order to keep the same colors like in Fig.2a and Fig.2b, the filling factor scale has been kept the same but offset by one as compared to Fig.2b. Therefore the boundaries of the $\nu_\uparrow = 1$ and the $\nu_\downarrow = 1$ layers are missing, while they add two more quantum channels to the system for achieving the $\nu = 6$ conductance plateau. In Fig.\ref{S1_density_nu6} it is also nicely seen, that in contrast to Fig.\ref{S2_density_nu5} the separation of the channels by the narrow almost missing bulk region is well developed as required for a well developed even $\nu = 6$ conductance plateau. 
\newline
\newline
{\bf APPENDIX D: Comparing our results with the results of Chklovskii, Shklovskii and Glazman}
\newline
\newline
In Fig.\ref{CSGvsHF} we compare our results with the results of Chklovskii, Shklovskii and Glazman\cite{Chklovskii1992} (CSG). Fig.\ref{CSGvsHF}a-f is the original figure of CSG and Fig.\ref{CSGvsHF}g is a reproduction of Fig.2b of the actual paper. For better comparison we flipped the x- and y - axis in Fig.\ref{CSGvsHF}g as compared to Fig.2b. Since CSG neglect spin-splitting, we use only the spin$\downarrow$ LLs for this comparison. On the left (Fig.\ref{CSGvsHF}a) there are two narrow edge channels as one would obtain from a non-interacting model. The narrow edge channels get widened to compressible stripes according to single particle interactions as proposed by CSG (Fig.\ref{CSGvsHF}d). As a guide for the eyes, arrows on top of the figure associate those stripes with the corresponding channels in Fig.\ref{CSGvsHF}a.  Following those arrows further to Fig.\ref{CSGvsHF}g on the right, the outer edge channel gets associated with an also quite narrow channel on the left of Fig.\ref{CSGvsHF}g. However, the inner wide edge stripe of Fig.\ref{CSGvsHF}d gets associated with a network of still narrow channels (green color in the middle of Fig.\ref{CSGvsHF}g) that covers a widened region. In Fig.\ref{CSGvsHF}g the widened compressible region with smoothly varying carrier density of Fig.\ref{CSGvsHF}f is replaced by a mixture of clusters of full end empty LLs, while the cluster boundaries act as narrow channels creating the channel network. The smoothly varying carrier density of Fig.\ref{CSGvsHF}f is only reproduced if turning-off the many body interactions by switching to pure Hartree calculation (compare Fig.\ref{S2_density}). 
\newline
\newline
\newline
{\bf APPENDIX E: Aharonov-Bohm oscillations of edge modes}
\newline
\newline
A most attractive manifestation of phase coherent quantum transport are quantum interference oscillations generated by edge modes. In this section we will discuss the typical demands for the observation of such oscillations. We use a recent study of Nakamura et al, where high precision Aharonov-Bohm (A-B) oscillations of quantum Hall edge modes have been achieved at fractional and integer filling\cite{Nakamura2019AharonovBohmModes}. From estimations concerning the geometrical demands for observing such oscillations it will become clear that these are possible just on the basis of narrow edge modes as delivered by our model only. 

The authors use a loop with an active area of $A=0.73 \mu m^2$ and achieve an oscillation period of $\Delta B=7 mT$ as a function of the magnetic field, while applying a magnetic field as high as $B=4.5 T$. This means that for changing the magnetic flux by exactly one flux quantum one can either change the magnetic field by $\Delta B=7mT$ or one could change the active area $A$ by a certain amount of $\Delta A$. The enclosed magnetic flux depends linearly on both quantities and the appropriate relation for those quantities therefore is $\Delta B/B=\Delta A/A$. From this relation $\Delta A$ appears to be $\Delta A=0.001 \mu m^2$ for creating one oscillation period. This means that a variation of the loop area by an amount of $\Delta A$ achieves a phase shift by $\Delta \phi = 2 \pi$. Consequently, any uncertainty of the area A must be well below that value of $\Delta A$ for keeping the phase uncertainty sufficiently below $\Delta \phi = 2 \pi$ for avoiding a smearing out of the oscillations because of multiple interference paths. For a square shaped structure of length $L=0.85 \mu m$ that value of $\Delta A$ transforms to a variation of the length by $\Delta L=0.66 nm$. In order to keep this precision of the length, also the geometrical width of the interference path would have to stay well below that limit. Like known from optical wave guides, such geometrical demands far below the wave length are physically not realistic and point clearly towards the requirement of mono-mode wave transmission for keeping the phase accuracy. It is well known, that the (transverse) modes of edge states are  associated with the cyclotron radius $R_C$ according to the particular Landau levels. By far, these limits cannot be achieved by the theory of CSG that produces almost macroscopically wide compressible stripes up to hundreds of $nm$ width. In contrast, our model meets this criterion by delivering exactly the required well defined mode structure of the channels, as can be seen in green color e.g. in Fig.\ref{S1_density_nu5} as well as for the next higher LL in Fig.\ref{S2_density_nu6}. In this context the channels in our simulation appear indeed as mono-mode electron wave guides with modes that are correctly associated with the cyclotron radius $R_C$ according to the corresponding LL.

\FloatBarrier
\bibliographystyle{prsty}
%\bibliography{SGM-QHE}

% end here if supplement is not included
\ifNOSUP\end{document}\else%

%%%%%%%%%%%%%%%%%%%%%%%%%%%%%%%%%%%%%%%%%%%%%%%%%%%%%%%%%%%%
%%
%% Supporting Information
%%
%% to be moved to its own file later
%%
%%%%%%%%%%%%%%%%%%%%%%%%%%%%%%%%%%%%%%%%%%%%%%%%%%%%%%%%%%%%
\clearpage\newpage
\setcounter{figure}{0}
\setcounter{table}{0}
\def\thefigure{S\arabic{figure}}
\def\thetable{S\arabic{table}}
\setcounter{page}{1}
\pagestyle{plain}
%%%%%%%%%%%%%%%%%%%%%%%%%%%%%%%%%%%%%%%%%%%%%%%%%%%%%%%%%%%%
%\section{Supplemental material}

\mbox{\bf SUPPLEMENTAL MATERIAL}
\newline
\newline
{\bf Revision of the edge channel picture of the integer quantum Hall effect}
\newline
\newline
{by J. Oswald}
\newline

%%%%%%%%%%%%%%%%%%%%%%%%%%%%%%%%%%%%%%%%%%%%%%%%%%%%%%%%%%%%%%%%%%%%

%%%%%%%%%%%%%%%%%%%%%%%%%%%%%%%%%%%%%%%%%%%%%%%%%%%%%%%%%%%%%%%%%%%%
\fi\end{document}